\documentclass[11pt,a4paper]{article}
\usepackage[hyperref]{eacl2021}
\usepackage{times}
\usepackage{colortbl}
\usepackage{tikz-qtree}
\usepackage{array,multirow}
\usepackage{makecell} 
\usepackage{amsmath}
\usepackage{amssymb}
\usepackage{float}
\usepackage{fancyhdr}
\usepackage{graphicx}
\usepackage{algpseudocode}
\usepackage{algorithm}
\usepackage{float}
\usepackage{bm}
\graphicspath{ {./images/} }
\algnewcommand\algorithmicswitch{\textbf{switch}}
\algnewcommand\algorithmiccase{\textbf{case}}
\algnewcommand\algorithmicassert{\texttt{assert}}
\algnewcommand\Assert[1]{\State \algorithmicassert(#1)}%
\algdef{SE}[SWITCH]{Switch}{EndSwitch}[1]{\algorithmicswitch\ #1\ \algorithmicdo}{\algorithmicend\ \algorithmicswitch}%
\algdef{SE}[CASE]{Case}{EndCase}[1]{\algorithmiccase\ #1}{\algorithmicend\ \algorithmiccase}%
\algtext*{EndSwitch}%
\algtext*{EndCase}%

\aclfinalcopy

\title{METL - a modern ETL pipeline with a dynamic mapping matrix}

\author{Christian Haase$^*$, Timo Röseler$^+$, Mattias Seidel$^+$\\
  $^+$ EOS Technological Solutions, Hamburg, Germany \\
  $^*$ Otto Business Intelligence, Hamburg, Germany\\
  \texttt{haase.mail@web.de} \\
\texttt{t.roeseler@eos-ts.com} \\
  \texttt{m.seidel@eos-ts.com} \\
  \\
  Preprint\\
\\}
\date{2022}
\begin{document}
\maketitle
\begin{abstract}
Modern ETL streaming pipelines extract data from various sources and forward it to multiple consumers, such as data warehouses (DW) and analytical systems that leverage machine learning (ML). However, the increasing number of systems that are connected to such pipelines requires new solutions for data integration. The canonical (or common) data model (CDM) offers such an integration. It is particular useful for integrating microservice systems into ETL pipelines.  \cite{Villaca:2020, Oliveira:2019} However, a mapping to a CDM is complex. \cite{Lemcke:2012} There are three complexity problems, namely the size of the required mapping matrix, the automation of updates of the matrix in response to changes in the extraction sources and the time efficiency of the mapping for near real-time use cases. 
\newline
In this paper, we present a new solution for these problems. More precisely, we present a new dynamic mapping matrix (DMM), which is based on permutation matrices that are obtained by block-partitioning and compacting the large and sparse mapping matrix. We show that the DMM can be used for automated updates in response to schema changes, for parallel computation in near real-time and for highly efficient compacting. For the solution, we draw on research into matrix partitioning and parallel computing \cite{Quinn:2004} and dynamic networks \cite{Haase:2021}. The DMM has been implemented into an app called Message ETL (METL). METL is the key part of a new ETL streaming pipeline at EOS that conducts the transformation to a CDM. The ETL pipeline is based on Kafka-streams. It extracts data from more than 80 microservices with log-based Change Data Capture (CDC) with Debezium and loads the data to a DW and an ML platform. EOS is part of the Otto-Group, the second-largest e-commerce provider in Europe.
\end{abstract}
\section{Introduction} 
The modernization of pipelines that extract, transform and load data (ETL) is a central aspect of the transformation of enterprise systems in the age of cloud computing and machine learning. \cite{Biswas:2020} The push for modernization  has been caused by the increased velocity, variety and volume of big data and the new demands of systems that utilize ML. \cite{McCarthy:2019} 
\newline
The modernization often encompasses the improvement of two different types of ETL pipelines. Traditional pipelines dating back to the 1990s connect a central database with a data warehouse (DW) and load the data once a day. They are too slow for modern near real-time scenarios and are often replaced with new streaming pipelines. These streaming pipelines are faster and can serve multiple systems more easily. They extract data from various sources, for example with CDC, and can forward it in near real-time to a multitude of consumers, such as ML systems. 
\newline
However, these modern streaming pipelines have also experienced new problems. With the increased number of connected distributed systems, the need has arisen to provide standardized data exchange formats for them. Such problems are in particular pressing for large microservice systems. Therefore, many enterprises have begun adding data integration to their ETL streaming pipelines. 
\newline
The canonical data model is a design pattern that offers such data integration. It was originally invented for service-oriented architectures (SOA). \cite{Forrester:2010} In recent years, the pattern has gained renewed attention for microservice systems \cite{Villaca:2020} and for distributed cloud systems \cite{Microsoft:2022}.  A custom CDM provides a bespoke alternative to the established commercial Kafka-streams platforms.
\newline
While a CDM offers many benefits, it also presents significant challenges and is difficult to implement. \cite{Lemcke:2012} Most mapping systems label large data packages with metadata, even in big data scenarios, such as railway systems. \cite{Suleykin:2020} A CDM, however, requires that every single attribute of the extraction schema is handled. 
\newline
In order to use a CDM efficiently in modern streaming ETL pipelines with near real-time requirements, more research and best practice examples are needed. We present such a pioneering study on the integration of a microservices system with a CDM into a streaming pipeline at EOS. EOS is part of the Otto-Group, which is the second largest e-commerce provider in Europe, and has specialized in financial services.
\newline
In the first part of this paper, we present and analyse the architecture of the pipeline and the required mapping system.
We show that the position of the CDM in modern ETL pipelines with microservices has changed as compared to older SOA-architectures. While a CDM used to be located between systems in an Enterprise Service Bus (ESB), it is now located inside the microservice system and acts more strongly as an API that decouples the microservice system from the enterprise platforms. We further analyse that the versioning of the extraction schemata of the various databases of the microservices,  forms the greatest challenge for the implementation of the mapping system with log-based Change Data Capture. We estimate that the mapping matrix for the EOS-pipeline can grow up to 1.000.000.000 elements, which in turn can grind the pipeline down. It needs to achieve near real-time performance for advanced ML-use cases.
\newline 
In the second part of the paper, we develop a static and sparse baseline system for the app, which is crucial to understand the mapping in a distributed system.  We show that METL is best conceptualized as a distributed dynamic network encompassing two tree-shaped schemata sub-graphs with blocks of attributes. The attributes describe data objects contained in the Kafka-messages. The data objects can either be mapped from one metadata system to another or they can be filtered out. This requires a mapping function with a parameter that can either have the value 1 for forwarding to another metadata system or 0 for filtering. The parameters for all possible connections between the attributes of the two metadata systems form the mapping matrix.
\newline
The baseline system exhibits three problems, namely the large size and sparsity of the matrix, the low time efficiency of the mappings and the inability to update the matrix in response to versioning changes of the schemata in the dynamic network. The greatest challenge of the system consists in the fact that the matrix needs to be updated. It expands when new blocks of attributes are added to the dynamic network or shrinks when blocks of attributes are deleted. We estimate that an update operation needs to calculate values for up to 100.000 new elements of the matrix.
\newline
In the third part of the paper, we present a new solution for automating updates to the matrix, for processing mappings with it in near real-time and for optimizing space efficiency. We are achieving this solution with a new dynamic mapping matrix that consists of a set of densely compacted permutation matrices. We are presenting two different strategies how such a DMM can be derived. The first one balances time- and space-efficiency. It partitions the overall matrix into blocks, deletes the null blocks and sizes the blocks down to the largest permutation matrix per block. It then partitions these permutation matrices and only saves the elements with the value 1 to a set. The super-set of these sets is the DMM.  This strategy is suitable for most systems. We further present an aggressive strategy that compacts the matrix even more, namely to unique square matrices by employing sequential pattern generalisation on version-super-blocks of the mapping blocks. It requires more computation for updates, but achieves an even higher storage efficiency. 
\newline
We show that both strategies achieves a compacting of more than 99\% for standard use cases and are providing the basis for optimal parallel computing with horizontal scaling and parallel execution of the mapping operations. The first strategy, however, offers a much easier update path for the matrix.
In the fourth and fifth part, we explain the implementation of the new app and the streaming pipeline and evaluate the performance of METL in the pipeline. 
\begin{figure*}[ht]
\includegraphics[width=\textwidth]{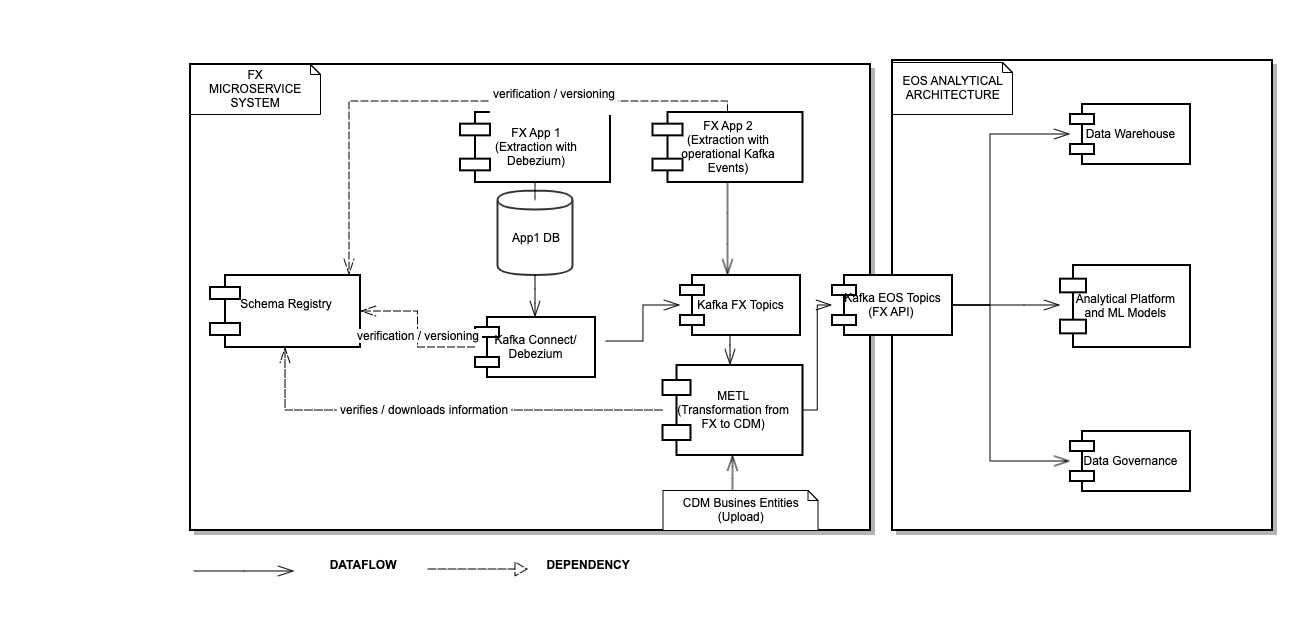}
\caption{\label{fig:architecture}Architecture of the modern streaming ETL pipeline that extracts data from more than 80 microservices, transforms it to a CDM with the app METL and then pushes the data via Kafka-streams to the data warehouse and machine learning systems. Note that the CDM mapping has been moved inside the microservice system as compared to older SOA-architecture where it was located between systems. The main problem of the architecture is the high number of extracting schema versions that are produced by the microservice system. They expand the mapping matrix to a very large size of up to 1.000.000.000 elements and trigger constant updates of it. This complexity can slow down the pipeline. The new dynamic mapping matrix solves this problem and enables a constant flow of data in the pipeline in near real-time.}
\end{figure*}
\section{Related research}
For the solution, we draw on research into dynamic networks \cite{Haase:2021, Rosetti:2018}, distributed systems, microservices, enterprise architecture, ETL pipelines, Kafka-streams, mapping systems and into calculations with very large matrices. \cite{Quinn:2004} 
\newline
Very large matrices play a role in many industrial, scientific and machine learning applications. The main approach for speeding up calculations with very large matrices is to partition them and to process the single parts in parallel. Such strategies are often developed in relation to static matrices, these are matrices that do not change their size or their values. There is less research on how to conduct such strategies in relation to dynamic matrices, which expand or shrink in size and undergo value updates.
\begin{figure*}[ht]
\begin{center}
\includegraphics[width=1\textwidth]{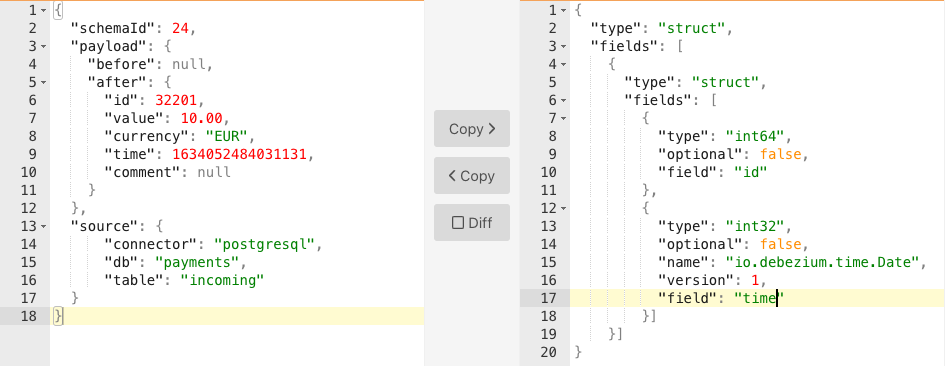}
\end{center}
\caption{\label{fig:schema} Left: example of a schematized Kafka-message. It transmits data that has been extracted by Debezium. Right: example of a (shortened) originating JSON-schema with a version that describes the content of the message.}
\end{figure*}
\section{Architecture of the new streaming ETL pipeline with a CDM}
The FX-system at EOS is a modern distributed microservice system that deals with financial operations and is driven by machine learning models that predict the next best action. One of these models predicts, for example, an optimal monthly rate for debt repayments. The modernization of the ETL pipelines at EOS has been caused by a number of issues that the previous pipelines and APIs have exhibited:
\begin{itemize}
    \item The previous ETL streaming pipeline between FX and the DW, thus a modern pipeline type, could not deal efficiently with the high change rate of the data structures in the microservice system.
    \item The ML models have been trained with data from a different system forwarded by a second ETL pipeline. This has led to skewed predictions for FX data.
    \item This second pipeline was an old ETL pipeline type, which was not fast enough for near real-time scenarios.
\end{itemize}
EOS decided to overcome these challenges with a new combined ETL streaming pipeline that contains a CDM inside the microservice system as a decoupling API. The task of the CDM is to filter out technical data from the microservices, condense partially duplicated data, reduce the high number of data structure changes to a more stable schema, offer more precise and unified descriptions of the business content of the data for the DW and the ML models and introduce more general data types for sharing. \cite{Forrester:2010} The new planned ETL pipeline shall have the potential for future near real-time cases. It is based on Kafka streams. It rests on five new pillars:
\begin{itemize}
\item Debezium connectors to the microservice databases that extract data in near real-time \cite{Mohrling:2022} , 
\item an Apicurio registry for the extracting schema which enforces rules for schema versioning and schema evolution,
\item a CDM for the outgoing messages that offers data integration,
\item a new mapping app called Message ETL (METL) that maps extracted data to the CDM and offers adaptations to version changes as well as the possibility for new initial loads and error management procedures,
\item a streaming API that hides the complexity of the distributed microservice system.
\end{itemize} 
The architecture of the new ETL streaming pipeline is shown in figure \ref{fig:architecture}. As mentioned above, the position of the CDM mapping constitutes a major architectural change as compared to the older SOA-architecture where the CDM was located between the single systems of an enterprise and not within one. METL conducts the mapping to the CDM and then pushes the data to the enterprise analytics system via Kafka-streams. The stream of schematized Kafka messages created by METL is the API of the microservice system to the enterprise systems.
\subsection{Kafka messages with data and metadata}
The data in the pipeline is forwarded with schematized Kafka messages. Figure \ref{fig:schema} shows a message (left) and a describing extracting schema that is used by Debezium (right).
The described data objects are called the payload of the message. The payload of the incoming message consists of pairs of schema attributes and data objects, such as "time":"1634052484031131". We also call the attributes a metadatum of the datum "1634052484031131". The attributes are part of the describing schema of the message.
\newline
The extraction schemata and CDM-schemata provide the same describing procedure. The CDM schema, however, differs from the extracting schema by offering more generalized types and better descriptions. Thus, for example, where the extraction schema on the right of figure \ref{fig:schema} contains the type "int32", the CDM schema would contain the type "integer". Furthermore, there are additional fields in the CDM schema for detailed short descriptions of the data, which are absent in the extracting schema shown in figure \ref{fig:schema}.
\newline
The mapping process changes the metadata, but not the data object itself.  The outgoing message can only contain data objects that were already present in the incoming one. "time":"1634052484031131", for example, could be mapped to "Time of the payment":"1634052484031131" in the outgoing message. The data object can also be filtered, this means it is not present in the outgoing message.
\newline
The best metaphor for understanding the mapping of payloads with abstract data objects is from transport networks related to a large shipping terminal. The mapping app that we design constitutes the harbour. The pipeline is the water flow. The Kafka-messages are the ships that dock to the harbour and contain a sealed load that consists of containers, i.e. the abstract data objects. It can also contain empty container spaces, which can be described by an attribute-data object pair of  "attribute":"null". Some of the existing containers are taken out of the incoming ship and are relabelled and are then put into new ships. There may also be empty container places in the new ships ("null"), but there cannot be two data containers at the same place. The outgoing ships carry these relabelled containers to their destinations.
\subsection{The two challenges: CDC and versioned schemata}
The pipeline forwards a high number of extracted data objects and uses two large and changing meta-data systems to describe these data objects which need to be converted to each other. This poses a triple challenge for the mapping app. 
\newline
The first challenge is the high number of data objects. This  is caused by the high number of data events in all databases that are extracted. Such an extraction CDC event is recorded in a special form, as shown in figure \ref{fig:schema}. The payload of a CDC event message consists of a "before" payload and an "after" payload. In the shown example, the "before" payload is empty. This means that a row has been created in a database. Since there was no row "before", the "before" payload is empty.  In addition, there are further CDC event types, such as changed metadata-events or deletion notifications.
\newline
The extraction of CDC events from a database happens in near real-time. It enables advanced near real-time use cases for systems that leverage machine learning. However, it requires a mapping system that can deal with frequent schema version updates. Each new column, deleted column, added table, deleted table, renamed column, or changed type in any database of any of the microservices, requires namely a new version of an extracting schema.
\newline
Having said that, it is important to note that even if one does not use CDC on database level for the extraction, one still needs to deal with a high number of events and versions in a large microservice system. The first ETL streaming pipeline of the FX system that extracted data from Kafka-based APIs only and not from the databases, also experienced a large amount of data structure changes.
\newline
Finally, the mapping system does not only need to deal with CDC events and changes to the extracting schemata, but also with the version updates of the CDM. Schemata of a CDM are called business entities. The focus on business content underlines the idea to filter out technical details in a CDM mapping. While the business entities are more stable than the extracting schemata, they also undergo changes.
\subsection{Semi-automated workflows in the ETL pipeline}
Among the three named challenges of the mapping system, the version updates of the extracting schema and the CDM schema pose the greatest challenges.
\newline
Version updates of the metadata systems affect all components of the pipeline, starting from the extraction to the CDM mappings to the final transformation and loading to the analytics systems. There are two main strategies for version updates, forward and backward compatibility. One allows the deletions of attributes, the other one additions. Such update paths are usually defined by an overarching schema-system for the pipeline, and exist for example also for the widely used Avro-schema \cite{Confluent:2022}.
\newline
Due to the knock-on effects of changes of the metadata systems for all subsequent stages of the pipeline, they need to be handled with care. Due to the high number of changes, the update process of the schemata is semi-automated. If the registry detects a change of attributes in a database, it enforces a semi-automated update process. The user needs to access the Apicurio schema registry, provide a new schema version that may only contain one single changed attribute, and then add new mapping rules to METL.  
\newline
Such an update of an extraction schema may require an additional update of a CDM schema. The update process of the schemata of the CDM, is done in a manual manner. There is numerous research in initializing and updating CDMs. \cite{Dietrich:2011, Dietrich:2013a, Dietrich:2013b}  In the EOS case, however, the data owners of the FX system took the opportunity to integrate their domain knowledge to the CDM and, therefore, update the business entities manually. 
\newline
Any update of an extraction schema or CDM schema requires also the adaption of the mapping matrix inside METL. For any new schema or CDM version, it expands and needs new values for the new matrix elements. For any deleted schema or CDM version, it shrinks and needs a deletion procedure.
\subsection{The mapping system as a distributed system}
The described elaborate version update process across the pipeline shall prevent technical and content-based mapping errors which can occur if an outdated mapping definition is used.  
\newline
However, there is a third source of possible complications that stems from the context of the implementation and the ETL pipeline. A mapping system that maps changing data structures in a pipeline is namely a distributed system and distributed systems produce problems of their own. They can be out of sync for example. It is for example possible that a new schema version has been pulled from the registry for a Kafka-message, but that this version is not known to METL yet.
\newline
Given the complexity of keeping a distributed mapping system in sync across the pipeline, it is good practice to have additional error-management procedures in place as well as options to set back Kafka-offsets and start new initial loads.
\newline
One needs to keep these fall-back options in mind when reading the paper. We conceptualize the overall mapping system with several components in the pipeline  with a state $i$. Each core element of this system, from the Kafka-messages to the schemata and the mapping matrix in METL inherits this state $i$. This does not mean that all elements can acquire a new state $i+1$ instantly. They need to be updated separately to a new state $i+1$ for the system to be in sync with an overall state again. In the implementation, we are thus checking  at several points if the METL app is in sync with the other components of the pipeline that contribute to the mapping and throw an error if this is not the case.
\subsection{Defining state change}
We can now define more precisely, how the mapping system changes from one state to another in relation to externally induced updates of the schemata. We define three triggers:
\begin{itemize}
\item The values of the mapping elements $^im_{qp}$ are changed by the user.
\item A version is added to the extraction schema or a version is deleted. This induces changes to the set of the schema attributes and requires changes in the number and possibly also the values of the mapping elements $^im_{qp}$.
\item A version is added to the CDM schema or a version is deleted. This induces changes to the set of all CDM attributes and requires changes in the number and possibly also the values of the elements $^im_{qp}$.
\end{itemize}
From this description, the main challenge of the mapping system becomes clearer. The size and the values of the mapping matrix need to be adapted to the external schema versions by transitioning the mapping matrix $^{i}M$ to $^{i+1}M$. However, what is the size $^{i}M$ and how many do we need to change in response to external updates? 
\newline
There are more than 10.000 data-attributes in total in all databases and Kafka-events of FX. The CDM has more than 1.000 attributes. We estimate that after some time of usage, there will be at least 10 versions of any schema that need to be managed in parallel. In order to conduct a mapping of all possible incoming Kafka message to the outgoing ones,  one needs up to 1.000.000.000 associative elements $^im_{qp}$ for any given state $i$. Thus for our $m$x$n$ matrix $^iM$ $ mxn \approx 1.000.000.000$.
\newline
We further estimate that any schema version consists of 10 attributes.  Thus, each time a new version is added to a database and a schema needs to be changed or each time a CDM-version is added, up to 100.000 elements $^im_{qp}$ need to be added, deleted or updated in the transition from $^{i}M$ to $^{i+1}M$. Such a very large number is difficult to handle. The diff-set is virtually impossible to update for a user without an automated procedure.
\subsection{Conclusion}
In this section, we have analysed the architecture and the challenges of constructing a mapping app in a distributed system. We have shown that the CDM has a unique position in the new microservice architecture. Furthermore, we have shown that version updates form a central part of the system and that they pose the greatest challenge for the new mapping app.
\section{Designing the basic mapping system}
In order to solve the challenges, we need to develop the mapping app in two steps. In the first step, that is described in this section of the paper, we are developing a baseline mapping system that is able to map data objects between Kafka messages in a distributed dynamic network. It can transform schematized Kafka-messages to a CDM, but it cannot update the matrix in an automated way yet. In the next section, we then solve all three problems of the mapping matrix of the baseline system, namely the size and sparsity of the matrix, the update-ability of the matrix and the time efficiency of the mappings with the matrix.
\subsection{The mapping system as a dynamic distributed network}
The main data structure of the mapping system is a distributed dynamic network. A network is a graph and a graph is a pair G = (V,E) with V a set of vertices and E a set of edges. In relation to networks, one often speaks of nodes and links instead of vertices and edges. Networks are often used to conceptualize real-live processes, such as data or traffic flows. Graphs are used for discussing more abstract concepts in mathematics.
\newline
The CDM tree and the tree of the extracting schema are two sub-graphs of the network. The extracting schema tree $^iD$ defines the domain of the mapping and  has a top-level node  $^id$. The CDM tree $^iR$ defines the range of the mappings and consists of a top-level node $^ir$.  These root-nodes have child-nodes in the form of the schemata. These are versioned. The versions define the attributes associated with any single business entity or the schema in this version.
\newline
We describe a path in such a sub-graph of the network with a simplified edge notation. Instead of defining a path as $^ir.be_1, be_1.v_1$ etc., we simply write $^ir$.$be_1$.$v_1$.$c_1$ (range.business-entity.version.cdm-attribute) or $^id.s_o.v_v.a_p$ (domain.schema.version.attribute). $c_q$ and $a_p$ are the attributes that we map.
\newline
We further define a number of sets in relation to the elements of the trees. All attributes $a_p$ and $c_q$ are elements of the sets $^iA$ and $^iC$. A specific schema version that encompasses a block of attributes is named as $^iD^o_v$ or $^iR^r_w$. Kafka messages contain pairs of attributes and data objects. The attributes that they contain correspond exactly to one versioned schema. We are thus also treating the pairs in the Kafka-messages as sets in relation to the schema trees and name the incoming and outgoing Kafka-messages $^iMIn^o_v$ and $^iMOut^r_w$.
\newline
A mapping is defined as an operation that associates each element of a given set, called the domain, with one or more elements of a second set, called the range. In our case, the domain and the range are constituted by the metadata trees and the associated data objects that are contained in the Kafka messages. Thus, our mapping function works on schema attributes and data objects that are best conceptualized as children of the attributes in a distributed and dynamic network.
\newline
In order for the Kafka message attributes to be related to a schema tree, both need to have the same state $i$. Once a Kafka-message is linked to the mapping network, it adds two child-nodes to the domain tree which contain values that are then used by the mapping function. 
\newline
For each of these connections, we define a parameter for a mapping function. The mapping function can either forward a data object and describe it with another attribute of the CDM metadata system or it can filter an object. It thus needs two parameter values, namely 0 or 1. All of these parameters form the $m$x$n$ mapping matrix $^iM$. The function multiplies the parameter value with the number of data objects that are described by one attribute of the extracting schema in one Kafka-message. Since a data object can only be "null" or non "null", this number of data objects can also only be 0 or 1. Therefore, the mapping function results either in a 0 or 1. If it results in 1, the data object contained in one of the child nodes is transferred and relabelled through the full network to an attribute in an outgoing Kafka-message. 
\newline
The child nodes of the associated attributes are always present in any Kafka-message, either implicitly or explicitly. The two child nodes are the data object $ad_p$ and the number of data objects $nad_p$. Since any abstract data object can only be "null" or non "null" in our JSON-based schemata, the number of abstract data objects per attribute is $nad_p  \in \{0,1\}$. The same two types of children are present in any outgoing Kafka-messages $MOut^r_w$, for which, $ncd_q \in \{0,1\}$. Since a "null" object is equivalent to saying that 0 data objects are described by a metadata-attribute, we formalize $\\ad_p = "null" \leftrightarrow nad_p = 0, \\ nad_p \neq "null" \leftrightarrow nad_p = 1,\\ cd_q = "null" \leftrightarrow ncd_q=0,\\ cd_q \neq "null" \leftrightarrow ncd_q = 1$
\newline
Figure \ref{fig:schema} shows a Kafka-message and its JSON schema. It contains, for example an  attribute with data object "time" : "1634052484031131".  We translate the information from the message as follows: Given p=1, then $a_1$ is "time", $ad_1$ is "1634052484031131" and $nad_p$ is 1. 
\begin{figure*}[ht]
\begin{center}
\begin{tabular}{| c |||c| c| c|||c| c||| c| c c c c }
\hline
& \cellcolor{magenta}{$^id.s_1$} & \cellcolor{magenta}{$^id.s_1$} & \cellcolor{magenta}{$^id.s_1$} & \cellcolor{magenta}{$^id.s_1$}  & \cellcolor{magenta}{$^id.s_1$} & \cellcolor{white}{$^id.s_2$}  \\
\hline
& \cellcolor{lightgray}{$v_1.a_1$} & \cellcolor{lightgray}{$v_1.a_2 $} & \cellcolor{lightgray}{$v_1.a_3$} & \cellcolor{pink}{$v_2.a_4\equiv v_1.a_1$}  & \cellcolor{pink}{$v_2.a_5\equiv v_1.a_3$} & \cellcolor{lightgray}{$v_1.a_6$}  \\
\hline
\hline
\cellcolor{lime}{$^ir.be_1.v_1.c_1$} &  \cellcolor{magenta}{$^im_{11}$} & \cellcolor{lime}{$^im_{12}$} & \cellcolor{magenta}{$\hdots$} & \cellcolor{lime}{$\hdots$} & \cellcolor{magenta}{$\hdots$} & \cellcolor{white}{$^im_{1^in}$}   \\
\hline
\cellcolor{lime}{$^ir.be_1.v_1.c_2$} & \cellcolor{lime}{$^im_{21}$} &\cellcolor{magenta}{$\hdots$} & \cellcolor{lime}{$\hdots$} & \cellcolor{magenta}{$\hdots$} & \cellcolor{lime}{$\hdots$} & \cellcolor{lime}{$\hdots$}  \\
\hline
\hline
\hline
\cellcolor{orange}{$^ir.be_1.v_2.c_3$} & \cellcolor{lightgray}{$\hdots$} & \cellcolor{orange}{$\hdots$} & \cellcolor{lightgray}{$\hdots$} & \cellcolor{pink}{$\hdots$} & \cellcolor{orange}{$\hdots$} & \cellcolor{lightgray}{$\hdots$}  \\
\hline
\cellcolor{orange}{$^ir.be_1.v_2.c_4$} & \cellcolor{orange}{$^im_{^im_1}$} &\cellcolor{lightgray}{$\hdots$} & \cellcolor{orange}{$\hdots$} & \cellcolor{orange}{$\hdots$} & \cellcolor{pink}{$\hdots$} & \cellcolor{orange}{$\hdots$}  \\
\hline
\end{tabular}
\end{center}
\caption{\label{fig:associationmatrix} The mapping system consists of a distributed network and a $^im$ x $^in$ matrix $^iM$ that contains $^im'$x$^in'$ sub-matrix mapping blocks. The blocks that are indicated with triple-lines provide the parameter values for the mapping function between any two Kafka-messages. The matrix is not only block-scoped, but also super-block-scoped due to the hierarchies in the schema-trees. Blocks (gray/orange, pink/orange) can be grouped into super-blocks of various versions related to one schema, for example (magenta/lime, white/lime). The paths are described with a short-notation in the form of node1.node2.node3. The equivalence sign $\equiv$ indicates that the same attributes have been assigned to multiple versions of the same schema within one magenta/white version-super-block. }
\end{figure*}
\subsection{Defining a single attribute mapping}
For every single mapping between two pairs of attributes and data objects, we define \textbf{a mapping element} $\bm{^im_{qp}}$ that defines the parameter of a mapping function. The value of this element can change across the states $i$. It can either be 0 or 1, $^im_{qp}  \in \{0,1\}$. As explained above, we define the data object as a child of the describing attribute. 
\newline
The mapping function takes its arguments from the value of the mapping element $^im_{qp}$ and from the children of the associated attributes $a_p$ and $c_q$, more precisely from the number of data objects $nad_p$ and $ncd_q$ that are described by the attributes. The mapping function works on the child nodes of $a_p$ and $c_q$. These are only present once we connect at least one incoming and one outgoing Kafka message with the state $i$ to this network. The mapping function is defined as: $c_q.ncd_q \gets ^im_{qp} * a_p.nad_p $ 
\newline
For brevity, we can say that we are mapping Kafka-messages. By this we mean that we are mapping pairs of attributes and data objects contained in such a message that are related to a larger schema tree. If the result of the function is $ncd_1 = 1$, then the data object from the incoming Kafka-message is attached to the attribute $c_q$ as a child element $cd_q$ in the outgoing Kafka-message.
\newline
JSON-based schemata can be constructed in such a way that "null" objects are omitted for the sake of brevity. However, this makes it more difficult to understand the mapping process. We thus, define that the Kafka-messages that the baseline system deals with, include all optional and "null" attributes. We, thus, define that all $a_p$ of any given $^iD^o_v$ are also present as descriptors in any $^iMIn^o_v$.
\subsection{Defining the mapping matrix}
In order to work with the large number of mapping parameters for all associations between all attributes, we aggregate the elements to a matrix $^iM$. All $^im_{qp}$ for one given state $i$ form the $m$x$n$ matrix  $^iM$. The sets $^i\mathcal{A}$ and $^i\mathcal{C}$ define the size of this matrix with $^im = |^i\mathcal{A}|$ and $^in = |^i\mathcal{C}|$.
\newline
The mapping matrix is block-scoped by the versioned schemata $^{i}D^o_{v}$ and $^{i}R^r_{w}$. We define $m'$x$n'$ sub-matrix blocks within this matrix. These mapping blocks map the data objects associated with the attributes of one versioned extracting schema to the attributes of one versioned CDM schema.
\newline
The matrix is also super-block scoped by hierarchies inherent in the schema trees. One can super-group attribute-blocks with different versions in relation to one particular schema o or r. This block- and super-block-scoped association matrix is shown in figure \ref{fig:associationmatrix}.
\newline
Mapping matrices are used in linear algebra for matrix-vector multiplication, thereby transforming a vector $x$ to a resulting vector $y$. However, such a matrix-vector system would be cumbersome to use and very sparse for our case. We are, thus, using the mapping matrix only as an ordered collection of associative elements with parameter values. We need to aggregate them to a matrix in order to conduct update operations on them and to obtain special blocks of elements that we can compact. Without a matrix, we could not conduct operations on the elements in an efficient manner and could not exploit the inherent block structures in the data. For the single mapping operations, we only use the single elements with the parameter value 1. We store these elements in sets.
\subsection{Names of matrix blocks}
For the definition of the various blocks in the matrix and the various sets of mapping elements that we obtain by partitioning the matrix, we need a naming scheme.
\begin{itemize}
\item SINGLE ELEMENT
\item $^im_{qp}$ - element of the Matrix $^iM$
\item SINGLE MATRIX BLOCKS AND SETS OF ELEMENTS
\item $MB$ - $m$x$n$ block of matrix elements related to one schema-version and one CDM-schema-version
\item $SB$ - square block of matrix elements - sub-matrix of $MB$
\item $NB$ - $1$x$1$ null block of matrix elements - sub-matrix of $MB$
\item $PM$ - largest permutation matrix of the block $MB$. It defines the 1:1 attribute mapping between two Kafka-messages.
\item $D$ - dense set of mapping elements, contains only elements with the value 1 that are obtained by partitioning a matrix block and discarding those elements that have the parameter value 0
\item SUPER-SETS OF MATRIX BLOCKS
\item $\mathcal{R}$ - Row super-block of matrix blocks
\item $\mathcal{C}$ - Column super-block of matrix blocks
\item $\mathcal{V}$ - Version super-block of matrix blocks
\item $\mathcal{U}$ - super-set with unique elements, i.e. $\mathcal{VUSB}$ - version super-block of unique square matrix blocks
\item SUPER-SETS OF SUPER-SETS
\item $^i\mathfrak{PM}$ - super-set of the largest permutation matrices per mapping block. It contains the column-, row- and version-block-sets of all permutation matrices.
\item $^i\mathfrak{USB}$ - super-set of all super-blocks of unique square block matrices; contains also all column-, row- and version-block-sets.
\item $^i\mathfrak{NB}$ - super-set of all square null blocks; contains also all column-, row- and version-sets  row sets.
\item $^i\mathfrak{MB}$ - super-set of all all $MB$; contains also all column-, row- and version-block-sets.
\end{itemize}
\subsection{The basic mapping algorithm}
The main approach of dealing with the large and sparse parameter matrix $^iM$, is to block-partition it. For this, we need  to group the elements $^im_{qp}$ of $^iM$ first into $^im'$x$^in'$ \textbf{blocks of mapping elements} $\bm{^i_{ov}MB_{rw}}$. Such blocks are indicated by triple lines in figure \ref{fig:associationmatrix}. Such a block contains all possible associative elements in relation to one $^iD^o_v$ and one $^iR^r_w$ for one given state $i$. This block can be used to map the attributes of one message to another.
\newline
We further define two constraints. First, we define that one incoming message is mapped by one single block to one outgoing message. One could design a system where an outgoing message is "filled up" by two subsequent incoming messages. However, we exclude this option explicitly. Second, we restrain the blocks to 1:1 attribute mappings. This prohibits awkward double-mappings of attributes and enables partitioning into independent elements with the value 1 for the dense sets that we use for parallel computation.
\newline
Based on these definitions and constraints, we can define the initial sparse and sequential mapping algorithm for one incoming Kafka message in Algorithm 1. The algorithm uses special column blocks of matrix mapping blocks. The algorithm creates an outgoing message with all possible data object pairs in the form attribute:"null". It only uses those elements that contain the value 1 for the mapping function, which multiplies the parameter 1 with the contained number of data objects per attribute. If the result is 1, the pre-constructed "null" object in the outgoing message is replaced with the associated mapped data object of the incoming message. The outgoing messages consists of messages that contain data objects and messages that contain only "null" objects. 
\newline
At the end of the paper, we replace this sequential and sparse algorithm with an optimized parallel algorithm that, first, only works on attributes in messages that are not "null", second, only works on parameter values that are not 0 without needing to filter them, third, only sends out messages that contain at least one non-"null" data object, and, fourth, works on all single mapping operations for any one incoming message in parallel.
\begin{algorithm}[h!]
\label{alg:OneMessageToOneMessage}
\caption{Sparse and sequential algorithm: mapping one $^iMIn^o_v$ to $^im'$ $^iMOut^r_w$}
\begin{algorithmic}[1]
\Procedure{Map}{$^iMIn^o_v$}
\State{get $^i\mathcal{CMB}^o_v$ from $^i\mathfrak{MB}$ that matches the indices of the incoming message}
\For{$\forall ^i_{ov}MB_{rw} \in ^i\mathcal{CMB}^o_v$}
\State $^iMOut^r_w \gets$ create message with pairs of all CDM-schema version attributes and "null" objects. Number of objects is implicit.$ \{\{^ir.be_r.v_w.c_1 = "labelCDM1", ^ir.be_r.v_w.c_1.cd_1 = \bm{"null"}, ^ir.be_r.v_w.c_1.ncd_1 = 0\} \hdots \}$
\For{$\forall$ $^im_{qp} \neq 0$ from the single element partition of $^i_{ov}MB_{rw}$}
\State for $^im_{qp}$ determine $^ir.be_r.v_w.c_q$ and $ ^id.s_o.v_v.a_p$
\State get child elements of $ ^id.s_o.v_v.a_p$ from $^iMIn^o_v$
\State $^ir.be_r.v_w.c_q.ncd_q \gets ^im_{qp} * ^id.s_o.v_v.a_p.nad_p $
\If{$ncd_q = 1$} 
\State{replace "null" object $\{^ir.be_r.v_w.c_1 = "labelCDM1", ^ir.be_r.v_w.c_1.cd_1 = \bm{"null"}, ^ir.be_r.v_w.c_1.ncd_1 = 0\}$ in $MOut^r_w$ with\\ $ \{^ir.be_r.v_w.c_1 = "labelCDM1", ^ir.be_r.v_w.c_1.cd_1 =$ $\bm{^id.s_o.v_v.a_1.ad_1}$, $^ir.be_r.v_w.c_1.ncd_1 = 1\}$ }
\EndIf
\EndFor
\State{collect all $^im'$ $^iMOut^r_w$}
\EndFor
\State{return $^im'$ $^iMOut^r_w$}
\EndProcedure
\end{algorithmic}
\end{algorithm}
\begin{figure*}[ht]
\begin{center}
\includegraphics[width=1\textwidth]{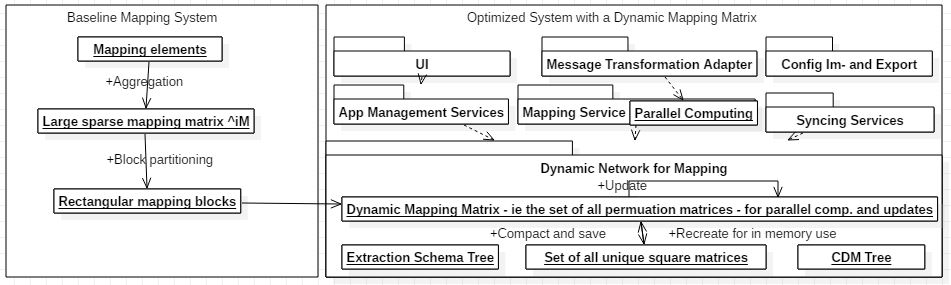}
\end{center}
\caption{\label{fig:dmmplan} Development of the DMM system: first, a static and sparse baseline mapping system is established with the mapping matrix $^iM$. For the initial sequential algorithm, a simple block partitioned form of the matrix is used. Second, an optimized dynamic system is developed. It uses a generalized pattern, namely the largest permutation matrices per mapping block. This set of permutation matrices is compacted to a dense set of single mapping elements. This set is the dynamic mapping matrix. It can be used for updates and parallel computation. Further, a different set of generalized square matrices can be derived for even more compact saving. This can be used as an alternative to the set of permutation matrices or in parallel for a hybrid implementation.}
\end{figure*}
\subsection{Conclusion}
The most important achievement of the baseline block-mapping system is that it defines the basic architecture of the mapping app in the pipeline as a distributed dynamic network with a mapping matrix. It can be used to transform one single Kafka-message from the stream to $^im'$ outgoing Kafka-messages with ease. The usage of $^im'$ sub-matrix blocks for the mapping is intuitive. However, overall the baseline system has three major flaws for our use case that are all related to the conception of the mapping matrix:
\begin{itemize}
 \item despite block-partitioning, it can only work on one single incoming Kafka-message at a time and can only map this one message to $^im'$ outgoing ones in a sequential manner. No conception for parallel computation with matrix blocks has been developed that is vital for near real-time use-cases;
    \item it cannot deal with updates of the mapping matrix yet that are triggered by updates of the trees in the dynamic network; 
   \item the space complexity of the sparse matrix and sparse matrix blocks is very high.
    \end{itemize}
\section{Developing the system with a dynamic mapping matrix}
The problems of the baseline system all relate to the large size and the static and sparse nature of the mapping matrix. We, thus, need to develop a strategy to reshape the matrix in such a way that the new form of the matrix is less sparse, can be used for updates and enables parallel computation.
\subsection{Reducing the CDM-schema versions}
We have shown that the very large size of the matrix and its expanding nature is mainly caused by the number of versions per schema. Naturally, the question arises, whether we need to keep all these versions in the system? The handling of parallel versions of the extracting schemata is necessary for testing different scenarios and updates in the various stages of the FX system. We cannot change this requirement. However, we do not need to keep multiple versions of the CDM-schemata. We are thus setting the new business rule for the dynamic system that outdated CDM-schema versions shall be deleted from the matrix regardless of the fact whether they are still used in the CDM-schema tree of the dynamic network. This reduces the maximal size of the matrix by the factor 10 from the estimated  1.000.000.000 elements to 100.000.000 elements.
\subsection{Main approach: matrix partitioning}
The main strategy for solving the size and sparsity problem of the matrix is to partition, compact and generalize it. Once we are dealing with sets of compacted and generalised matrix blocks, we have a much more flexible and dynamic data structure that enables us to solve the three named problems.
\newline
There are various strategies for partitioning and compacting very large matrices for parallel computation. \cite{Quinn:2004} We present two separate strategies.
 The first one is centred on generalising matrix blocks into permutation matrices and deleting all null blocks.  The strategy aims to achieve a balanced optimisation of time and space efficiency. The resulting matrices form the set $^i\mathfrak{PM}$ which we can use for parallel computing and updates. This strategy achieves a compacting of an estimated 99.9\% for our use case and enables highly efficient calculations. It constitutes a general solution strategy that is suitable for most mapping use cases that require near real-time performance but do not need to prioritise space efficiency as the top goal. 
\newline
The second strategy pursues a much more aggressive compacting strategy for optimizing space efficiency only. It takes the version-super-blocks of the matrix as a starting point, deletes all blocks that contain only nulls and then reduces the mapping blocks within the remaining ones to unique square matrices, consisting of unique permutation matrices and special null blocks. These form the set of unique square matrices $^i\mathfrak{USB}$. We can use this strategy for storing the matrix in a database in use cases that require highly optimized storage. For updates to the matrix and for parallel computation, we need to decompact the set, which is costly. This strategy achieves a higher compacting ratio than the first strategy in our CDM-mapping use case.
\newline
The DMM system that we develop in this sections, thus, can be realised by two different data sets, first the set $^i\mathfrak{PM}$, which is the set of all permutation matrices, and, second, the set $^i\mathfrak{USB}$ that is the set of all unique permutation matrices and special null blocks. The crucial sets for parallel computation and storage are the dense sets $^i\mathfrak{DPM}$ and $^i\mathfrak{DUSB}$ that result from either of the two strategies. They contain only elements with the value 1 that are obtained by partitioning a matrix block and discarding those elements that have the parameter value 0. 
\newline
Although the set $^i\mathfrak{DUSB}$ is more strongly compacted than $^i\mathfrak{DPM}$, both sets provides very strong null compacting. Both strategies compact a null matrix to null elements. The size of both sets $^i\mathfrak{DPM}$ and $^i\mathfrak{DUSB}$ grows in the worst case in the same linear way in relation to the number of realised mappings. The space complexity for the execution of one single mapping to $^im'$ outgoing message with the DMM is thus O(n), regardless of the chosen strategy.
\newline
All partitioning and compacting strategies for sparse matrices need to be applied with the matrix shape and the data content in mind to find an optimal solution that optimises the space efficiency for an average use case. The first strategy can be applied to any mapping matrix. The aggressive strategy, however, makes more use of the peculiar shape and value patterns of the mapping matrix for our specific use case of integrating a CDM into a microservice system. The estimated row:column ratio of the mapping matrix is 1:100. Therefore, the aggressive strategy runs a horizontal compacting algorithm on the CDM-schema versions in the matrix.
\newline
The set $^i\mathfrak{DPM}$ that we derive by the first strategy is at the centre of the DMM system that we present in this paper. It contains the densely saved permutation matrices, which enable parallel computation and updates of the matrix $^iM$ in an efficient manner for most use cases. It is thus, the "dynamic mapping matrix" in our case.  However, the set $^i\mathfrak{DUSB}$ can also play the role as the "dynamic mapping matrix". It is suitable for CDM-mapping systems that require more compacting and have less stringent near real-time requirements. Alternatively, both sets can also be implemented in parallel for a highly optimized hybrid strategy. We will be using this hybrid strategy for our implementation.
\subsection{Deriving the compacted dynamic mapping matrix}
\begin{figure*}[ht]
\begin{center}
\begin{tabular}{| c |||c| c| c|||c| c||| c| c c c c }
\hline
& \cellcolor{magenta}{$s_1.v_1$} & \cellcolor{magenta}{$s_1.v_1$} & \cellcolor{magenta}{$s_1.v_1$} & \cellcolor{magenta}{$s_1.v2$}  & \cellcolor{magenta}{$s_1.v_2$} & \cellcolor{white}{$s_2.v_1$}  \\
\hline
& \cellcolor{lightgray}{$a_1$} & \cellcolor{lightgray}{$a_2 $} & \cellcolor{lightgray}{$a_3$} & \cellcolor{pink}{$a_4\equiv a_1$}  & \cellcolor{pink}{$a_5\equiv a_3$} & \cellcolor{lightgray}{$a_6$}  \\

\hline
\hline
\cellcolor{orange}{$be_1.v_2.c_3$} & \cellcolor{green}{1} & \cellcolor{yellow}{0} & \cellcolor{brown}{0} & \cellcolor{cyan}{1} & \cellcolor{brown}{0} & \cellcolor{red}{0}  \\
\hline
\cellcolor{orange}{$be_1.v_2.c_4$} & \cellcolor{brown}{0} &\cellcolor{yellow}{0} & \cellcolor{green}{1} & \cellcolor{brown}{0} & \cellcolor{cyan}{1} & \cellcolor{yellow}{0}  \\

\hline
\hline
\hline
\cellcolor{lime}{$be_2.v_1.c_5$} & \cellcolor{red}{0} & \cellcolor{yellow}{0} & \cellcolor{yellow}{0} & \cellcolor{red}{0} & \cellcolor{yellow}{0} & \cellcolor{green}{1}   \\
\hline
\hline
\hline
\cellcolor{orange}{$be_3.v_1.c_6$} & \cellcolor{brown}{0} & \cellcolor{green}{1} & \cellcolor{yellow}{0} & \cellcolor{green}{0} & \cellcolor{yellow}{0} & \cellcolor{red}{0}\\
\hline
\cellcolor{orange}{$be_3.v_1.c_7$} & \cellcolor{green}{1} &\cellcolor{brown}{0} & \cellcolor{yellow}{0} & \cellcolor{yellow}{0} & \cellcolor{yellow}{0} & \cellcolor{yellow}{0} \\
\hline
\end{tabular}
\end{center}
\caption{\label{fig:dmm} Creating the dynamic mapping matrix, i.e. the dense sets $^{i}\mathfrak{DPM}$ and $^{i}\mathfrak{DUSB}$, from the sparse matrix $^iM$: We are presenting two algorithms that partition and compact the matrix $^iM$ for the creation of the DMM. The first algorithm, partitions the matrix into single rectangular block-matrices (with triple line borders). It then deletes the null blocks. The remaining blocks are sized down to square matrices by sub-matrix formation. This process obtains largest permutation matrices per block (green/brown and blue/brown). 
The permutation matrices are then block partitioned. All elements with the value 1 (green and blue) are saved to $^{i+1}\mathfrak{DPM}$. The efficient standard algorithm \ref{alg:transformtopm} compacts the above matrix from 30 to 7 elements. The second algorithm, partitions the matrix into version-super-blocks (magenta/white). It then deletes the version blocks that only contain zeros (red). It then sizes the mapping blocks in the version blocks down to square matrices, namely the largest permutation matrices and special 1x1 null-blocks (green). The green values are compacted and saved without 0 values to the set $^{i+1}\mathfrak{DUSB}$. The aggressive algorithm \ref{alg:transformtodmm} compacts the above matrix from 30 to 5 elements with a special 6th element. For the given use case, both algorithms achieve a compaction rate of more than 99\%.  }
\end{figure*}
\subsubsection{The balanced strategy: Creating the set $^i\mathfrak{DPM}$}
The first strategy begins with the partitioning of the matrix $^iM$ into rectangular mapping blocks ${^i_{ov}MB_{rw}}$. Each of these mapping blocks maps the elements of one sparse Kafka-message vector to another sparse Kafka-message vector. However, instead of using the mapping blocks directly for the mapping algorithm as in the baseline system, we now first delete all null mapping blocks from the result set. They are creating messages that only contain "null" objects. We delete these, because we do not need these messages. We are, thus, only interested in those mapping blocks that contain at least one mapping element with the value 1. As mentioned above, we usually only find one mapping block with at least one 1 for a single incoming Kafka-message. Since we have c. 100 possible mapping blocks for each incoming message, this step already compacts the matrix by an estimated 99\% for our use case. 
\newline
In the second step, we generalize the remaining mapping blocks to the largest permutation matrices that we can find per block. The largest permutation matrices per mapping block contain exactly all those 1 and 0 values that define a 1:1 attribute mapping between any two Kafka-messages. These permutation matrices are colored green/brown and blue/brown in figure \ref{fig:dmm}. We are thus deleting columns and rows that contain only 0s from the mapping blocks with a 1.
\newline
The key for optimizing the system with strategy one is the relationship between the rectangular matrix blocks and the permutation sub-matrices. If a matrix block has at least one 1, it contains a largest permutation matrix. It follows: \\ 1:1 mapping of attributes in a $h$x$j$ block of attributes with at least one 1 $\leftrightarrow$ a largest $k$x$k$ permutation sub-matrix of the $h$x$j$ block exists.
\newline
Finally, we block-partition the obtained largest permutation matrices into single elements and delete those elements that have a parameter value of 0. We expect that each block maps c. 10 attributes to 10 other attributes. Since there can be only up to 10 elements with the value 1, this step compacts the matrix further by an estimated 99.9\% in our use case.
\newline
Since the permutation matrices have linear independent rows and columns, we have further proof that the obtained elements with the value 1 are independent and can be used for parallel computing. This dense set of the block-partitioned largest square permutation matrices per mapping block is called $^i\mathfrak{DPM}$.
\newline
We can create the set $^i\mathfrak{DPM}$ in different ways. It is possible, for example, to use the User Interface or a CSV upload. However, the described way of partitioning and generalising the overall parameter matrix is the main strategy. The transformation is crucial for the optimized system. We will later need it to conduct it after each update. The algorithm \ref{alg:transformtopm} is shown below. 
\begin{algorithm}[h]
\caption{Transform $^iM$ to $^i\mathfrak{DPM}$}
\begin{algorithmic}[1]
\Procedure{Transform}{$^iM$}
\State{$^i\mathfrak{DPM}$ = \{\}}
\State{partition $^iM$ into $^i_{ov}\mathcal{MB}_{rw}$ }
\For{$\forall ^i_{ov}\mathcal{MB}_{rw} \neq 0 \in ^iM$}
\State{$^i_{ov}PM_{rw} \gets$ largest permutation matrix in $^i_{ov}MB_{rw}$}
\State{$^i_{ov}DPM_{rw} \gets$ block-partition $^i_{ov}PM_{rw}$ and delete all elements with value 0  }
\State{$^i\mathfrak{DPM} \cup ^i_{ov}\mathcal{DPM}_{rw}$}
\EndFor
\State{return $^i\mathfrak{DPM}$}
\EndProcedure
\end{algorithmic}
\label{alg:transformtopm}
\end{algorithm}
\subsubsection{The aggressive strategy: creating the set $^i\mathfrak{DUSB}$}
For the aggressive compacting strategy, we need a different starting point, namely version-super-blocks of the versions of the extracting schemata. In a first step, we thus, partition the matrix  $^iM$ into all version-super-blocks $^i_{o}\mathcal{VMB}_{rw}$.
\newline
In a second step, we then delete all version-super-blocks that contain only null mapping blocks. Comparable to the first step, we are reducing the matrix by 99\% by this null deletion strategy.
\newline
In a third step, we reduce the mapping blocks $^i_{ov}{MB}_{rw}$ in each of the remaining version-super-block to the largest square permutation matrices and \textbf{special square 1x1 null blocks} $\bm{^i_{ov}NB_{rw}}$. One $NB$ contains just one single element. These 1x1 square sub-matrices of the rectangular block matrices are the single green 0s or the single red 0s in figure \ref{fig:dmm}.
\newline
We then run a sequential pattern recognition algorithm on this set of square matrices per version-super-block. This algorithm determines unique square blocks in a sequence from the lowest to the highest version and deletes all other ones. It excludes null blocks in the lowest version. The unique permutation matrices are colored green/brown in figure \ref{fig:dmm}. If a $NB$ is encountered in the lowest version, it is ignored. Such null blocks are colored red in figure \ref{fig:dmm}. This second step of the second strategy is quite efficient and compacts the overall matrix further down to an estimated c. 99.5\% for our use case.
\newline
After the algorithm has found all unique square patterns, it compacts the set $^i\mathfrak{USB}$ to the dense set $^i\mathfrak{DUSB}$. The compacting algorithm processes the permutation matrices as explained above.  The \textbf{dense null-block} $\bm{^i_{ov}DNB_{rw}}$ is a special case among the dense blocks, because it is empty. This is realised in our implementation with the help of a hierarchical object structure, in which we save blocks with their indices as the top level object and the mapping elements as associated lower level elements. A block without mapping elements is a special null block. 
\newline
This last compacting step reduces the number of elements further to an estimated compaction rate of more than 99.9\% for the aggressive strategy in our use case.
We do not save another block, namely the \textbf{non-saved special null block}. This peculiar block only exists tucked away in the compacting algorithm of the $^i\mathfrak{USB}$. The set $^i\mathfrak{USB}$ is used to reconstruct a matrix based on an algorithm that replaces values in a null matrix in a sequential manner starting from the lowest version. Therefore, we can omit special null blocks from this set that start a sequence in the lowest version of a schema. The algorithm \ref{alg:transformtodmm} transforms the mapping matrix to $^i\mathfrak{DUSB}$. It is shown below.
\begin{algorithm}[h]
\caption{Transform $^iM$ to $^i\mathfrak{DUSB}$}
\begin{algorithmic}[1]
\Procedure{Transform}{$^iM$}
\State{$^i\mathfrak{DUSB}$ = \{\}}
\State{partition $^iM$ into $^i_{o}\mathcal{VMB}_{rw}$ }
\For{$\forall ^i_{o}\mathcal{VMB}_{rw} \in ^iM$}
\State{$ ^i_{o}VUSB_{rw}$ = \{\}}
\For{$\forall ^i_{ov}\mathcal{MB}_{rw} \in ^i_{o}\mathcal{VMB}_{rw}$ in ascending order of v}
\If{$^i_{ov}\mathcal{MB}_{rw} \neq 0$}
\State{$^i_{ov}SB_{rw} \gets$ largest square permutation matrix in $^i_{ov}MB_{rw}$}
\If{the latest $^i_{ov}SB'_{rw}$ that has been added to $ ^i_{o}VUSB_{rw}$ is not equivalent to $^i_{ov}SB_{rw}$ or  $^i_{o}VUSB_{rw} = \{\}$}
\State{$^i_{o}VUSB_{rw} \cup ^i_{ov}SB_{rw}$}
\EndIf
\EndIf
\If{$^i_{ov}SB_{rw}$ type of NB}  
\If{the latest $^i_{ov}SB'_{rw} $ that has been added to $\in ^i_{o}VUSB_{rw}$ is type of PM and $^i_{o}VUSB_{rw} \neq \{\}$ }
\State{$^i_{o}VUSB_{rw} \cup ^i_{ov}SB_{rw}$}
\EndIf
\EndIf
\EndFor
\State {$^i_{o}DVUSB_{rw} \gets$ partition and compact all $USB \in ^i_{o}VUSB_{rw}$ using a special format for the "empty" null block}
\State{$^i\mathfrak{DUSB} \cup ^i_{o}DVUSB_{rw}$}
\EndFor
\State{return $^i\mathfrak{DUSB}$}
\EndProcedure
\end{algorithmic}
\label{alg:transformtodmm}
\end{algorithm}
\subsubsection{Decompacting $^i\mathfrak{DPM}$ to $^iM$}
For various tasks in the DMM system, we need to decompact $^i\mathfrak{DPM}$ or $^i\mathfrak{DUSB}$ to $^iM$. The decompacting of $^i\mathfrak{DPM}$ is simple. We create a $^im$x$^in$ null matrix $^iM$ with $^im = |^i\mathcal{A}|$ and $^in = |^i\mathcal{C}|$ and then replace the values of those elements that are stored in $^i\mathfrak{DPM}$ with a 1.
\newline
The decompacting of $^i\mathfrak{DUSB}$ is more complicated. We first decompact $^i\mathfrak{DUSB}$ to $^i\mathfrak{USB}$, then create a null matrix $^iM$ and then reassign the stored unique blocks in a sequential manner until the full matrix is recreated. 
\begin{algorithm}[ht]
\caption{Decompacting $^i\mathfrak{DUSB}$ to $^iM$}
\begin{algorithmic}[1]
\Procedure{Decompacting}{$^i\mathfrak{DUSB}$}
\State{$m = |^{i}\mathcal{C}|$, $n = |^{i}\mathcal{A}|$}
\State{initialise null matrix $^{i}M$}
\State{initialise all blocks $^{i}_{o}{VMB}_{rw}$ and $^{i}_{ov}MB_{rw}$ in $^{i}M$}
\State{decompact $^i\mathfrak{DUSB}$ to $^i\mathfrak{USB}$ by creating the smallest possible square matrix, thus at minimum a 1x1 matrix, from the stored values, adding 0 values if necessary}
\For{$^i_{o}{VUSB}_{rw} \in ^i\mathfrak{USB}$}
\For{$^i_{ov}SB_{rw} \in ^i_{o}{VUSB}_{rw}$ for ascending v}
\If{there is a next element in $^i_{o}{VUSB}_{rw}$ }
\State{$v2 \gets $ version of next element}
\Else
\State{$v2 \gets$ \{get the version super-block $^{i}_{o}{VMB}_{rw}$ from $^iM$ with the same o,r,w as $^i_{o}{VUSB}_{rw}$ and choose $^i_{ov}{MB}_{rw}$ with highest v within it; return v\}}
\EndIf
\State{a = 0}
\While{a + v $\neq$ v2 }
\State{$^{i}_{o{v+a}}MB_{rw} \gets$ $ ^i_{ov}SB_{rw}$}
\State{a = a+1}
\EndWhile
\EndFor
\EndFor
\State{return $^{i}M$}
\EndProcedure
\end{algorithmic}
\label{alg:createpmfromdmm}
\end{algorithm}
\newline

\begin{figure*}[ht]
\begin{center}
\begin{tabular}{| c| c |||c| c| c|||c| c| c|||c| c| c c c  }
\hline
&& \cellcolor{magenta}{$s_1v_1$} & \cellcolor{magenta}{$s_1v_1$} & \cellcolor{magenta}{$s_1v_1$} & \cellcolor{magenta}{$s_1v_2$}  & \cellcolor{magenta}{$s_1v_2$ } & \cellcolor{magenta}{$s_1v_2$} &
\cellcolor{magenta}{$s_1v_3$}  \\
\hline
&& \cellcolor{lightgray}{$a_1$} & \cellcolor{lightgray}{$a_2$} & \cellcolor{lightgray}{$a_3$} & \cellcolor{pink}{$a_4 \equiv a_1$}  & \cellcolor{pink}{$a_5$ } & \cellcolor{pink}{$a_6 \equiv a_2$} &
\cellcolor{lightgray}{$a_7\equiv a_4$} \\
\hline
\hline
\hline
\cellcolor{magenta}{$s_1.v_1$} &  \cellcolor{pink}{$c_1$} & \cellcolor{red}{1} & \cellcolor{red}{0} & \cellcolor{red}{0} & \cellcolor{red}{1} & \cellcolor{red}{0} & \cellcolor{red}{0} & \cellcolor{red}{1}  \\
\hline
\cellcolor{magenta}{$s_1.v_1$} &  \cellcolor{pink}{$c_2$} & \cellcolor{red}{0} &\cellcolor{red}{0} & \cellcolor{red}{1} & \cellcolor{red}{0} & \cellcolor{red}{0} & \cellcolor{red}{1} & \cellcolor{red}{0}  \\
\hline
\hline
\hline
\cellcolor{magenta}{$s_1.v_2$} &\cellcolor{lightgray}{$c_3 \equiv c_1$} & \cellcolor{green}{1} & \cellcolor{yellow}{0} & \cellcolor{brown}{0} & \cellcolor{green}{1} & \cellcolor{yellow}{0} & \cellcolor{brown}{0} & \cellcolor{green}{1} \\
\hline
\cellcolor{magenta}{$s_1.v_2$} &\cellcolor{lightgray}{$c_4 \equiv c_2$} & \cellcolor{brown}{0} &\cellcolor{yellow}{0} & \cellcolor{green}{1} & \cellcolor{brown}{0} & \cellcolor{yellow}{0} & \cellcolor{green}{1} & \cellcolor{yellow}{0}  \\
\hline
\hline
\hline
\cellcolor{white}{$s_2v_1$} &\cellcolor{pink}{$c_6$} & \cellcolor{brown}{0} & \cellcolor{green}{1} & \cellcolor{yellow}{0} & \cellcolor{yellow}{0} & \cellcolor{yellow}{0} & \cellcolor{yellow}{0} & \cellcolor{yellow}{0}  \\
\hline
\cellcolor{white}{$s_2v_1$} &\cellcolor{pink}{$c_7$} & \cellcolor{green}{1} &\cellcolor{brown}{0} & \cellcolor{yellow}{0} & \cellcolor{yellow}{0} & \cellcolor{yellow}{0} & \cellcolor{yellow}{0} & \cellcolor{yellow}{0} \\
\hline
\end{tabular}
\end{center}
\caption{\label{fig:dmm2} Updating the dynamic mapping matrix $^i\mathfrak{DPM}$ from state $^i$ to state $^{i+1}$: The algorithm works on sets only, but is demonstrated here in relation to the full matrix. Given two external events, namely (1) the addition of a new extracting schema version $s_1.v_3$ and (2) the addition of a new business entity version $s_1.v_2$ with attributes $c_3$ and $c_4$. For each event the automated update is triggered.  For event (1), the system creates a new column block with 0 values (yellow) and partitions it. It then copies the 1 values of the permutation matrices in the $^i\mathfrak{DPM}$ to a matching new null mapping block for those columns that map equivalent attributes. In this process, null blocks, new equivalent permutation matrices or new smaller ones can be created. It then adds the  new permutation matrices  (green/brown) to $^i\mathfrak{DPM}$. Finally, the system informs the user about newly created smaller permutation matrices. The process for event (2) is similar, but works on row-level and has one additional step. After the update the old version of the CDM-schema is deleted (red).}
\end{figure*}
\subsection{Updating the DMM}
We have shown that the dynamic mapping matrix reduces the size of the mapping matrix. The second great benefit consists in the fact that it can be used for effective updates with minimal overhead. We need to update the DMM in response to four triggers, namely the addition of a new version or the deletion of a version of any of the two schemata. 
\subsubsection{Attribute equivalences as the basis for automated updates}
For the addition of versions we need to create new rows and columns in the mapping matrix and assign new values to it. Our main strategy to finding new values consists in the copying of known values. This relies on the observation that equivalences among attributes across versions occur often because attributes are duplicated across schema versions in the update process. If we have a version 1 with attributes $a_1$ and $a_2$ and we add $a_3$, then version 2 consists of $a_4 \equiv a_1$, $a_5 \equiv a_2 $ and $a_3$. Subsequently, we can copy the known values for $a_1$ and $a_2$ to $a_4$ and $a_5$.
\newline
This duplication of attributes is one of the main reasons for the very quick and large expansion of the mapping matrix in a microservice system with a CDM. However, it also provides the basis for the solution. In more abstract terms the solution consists of the generalisation of the attributes per schema across versions.
\subsubsection{Manual and semi-automated updates}
In order to conceptualize the matrix updates in more detail, we need to define which steps we can automate and which steps require user intervention. First, for the majority of cases we can apply an automated update procedure without any user intervention. This is the case for all deletions of versions from the network and for additions of new versions that do not change mapped attributes. In such a case we can either simply delete column- or row-sets from the DMM or we can add new matrix elements and create new values by copying known values.
\newline
Second, there are many cases for semi-automated workflows. These are automated updates of the matrix in response to additions of schemata versions which may require a confirmation or additional update by the user. The standard case for such a semi-automated workflow consists in the deletion of a mapped attribute, which requires an added schema version. This triggers an automated addition update of the matrix that copies known values. This results in a new and smaller permutation matrix or a new null block because we cannot reassign a known value. In both cases, the user should double-check if further updates of the new matrix block are needed.
\newline
Finally, there are two cases that require the setting of the values by a user, namely when the first version of a schema is added and when a single matrix block needs to be updated. Both cases can be realised via an User Interface. The initialisation can also be done via an upload of a CSV file. In sum, it can be said that automated updates should form the core of the DMM system. These need to be paralleled by an API for initial uploads and an User Interface.
\subsubsection{Automated updates}
The automated updating algorithm on the full matrix works on the rows and columns of the association matrix $^{i}M$. For a column or row deletion in the matrix that is triggered by a version deletion in a schema, for example, we simply delete the blocks of columns or rows from $^iM$ that associate these attributes. We then derive $^{i+1}M$.
\newline
We translate these matrix-based operations to set-based operations on our super-super-set $^i\mathfrak{DPM}$. This enables us to save a lot of overhead. The deletions can be easily executed. For the addition of a new set $^{i+1}D^o_{v+1}$ to the schema tree, we iterate over the sets $^i_{ov}DPM_{rw}$ in a column super-set $^i_{ov}\mathcal{DCPM}$ in $^i\mathfrak{DPM}$ in relation to the previous version v. The single mapping sets map elements $a_p$ from $^{i+1}D^o_{v}$. Second, we look up equivalent values $a_p \equiv a_{p'}$ in $^{i+1}D^o_{v+1}$. For each equivalent value we create a new mapping element and add it to a new dense set that we add to $^i\mathfrak{DPM}$ to create $^{i+1}\mathfrak{DPM}$. In this process, we may create new smaller permutation matrix as compared to version v if we cannot reassign all values of a block of version v to a new block in version v+1. Finally, we inform the user about such newly created permutation matrices.
\newline
After such an update process, we also need to clean up. In our case, we only clean up for additions of CDM-schemata versions as we explicitly want to use different extracting schemata versions in parallel and do not want to delete them. We have set the rule that we only want to map any version of an extracting schema to one business entity version. This rule means that we need to delete old CDM-version blocks of rows after a vertical update operation from the matrix.
\subsubsection{The algorithm for updating $^i\mathfrak{DPM}$ to $^{i+1}\mathfrak{DPM}$}
We design the automated algorithm for matrix updates with the above mentioned four triggers in mind. They determine the overall structure of the algorithm as a switch statement with four cases. The algorithm \ref{alg:updatefromdmm} is automatically triggered when a schema changes. It is shown on the right.
\begin{algorithm}[h!]
\caption{Update set $^{i}\mathfrak{DPM}$ to $^{i+1}\mathfrak{DPM}$ in response to schema version additions or deletions}
\begin{algorithmic}[1]
\Procedure{Auto-Update}{$^{i}\mathfrak{DPM}$, change case: (1) deleted $^iD^o_v$, (2) deleted $^{i}R^r_{w}$, (3) added $^{i+1}D^o_v$, (4) added $^{i+1}R^r_w$}
\Switch{change-case}
\Case{(1) deleted $^iD^o_v$}
\State{$^{i+1}\mathfrak{DPM} \gets ^{i}\mathfrak{DPM}$ $\setminus$ all $^i_{ov}DPM_{rw} \in ^{i}\mathfrak{DPM}$ with the same o and v as $^iD^o_v$}
\EndCase
\Case{(2) deleted $^{i}R^r_{w}$}
\State{$^{i+1}\mathfrak{DPM} \gets ^{i}\mathfrak{DPM}$ $\setminus$ all $^i_{ov}DPM_{rw} \in ^{i}\mathfrak{DPM}$ with the same r and w as $^iR^r_w$}
\EndCase
\Case{(3) added $^{i+1}D^o_{v+1}$}
\State{$^{i+1}\mathfrak{DPM} \gets^{i}\mathfrak{DPM}$}
\For{for all mapping sets of previous version: $\forall ^i_{ov}DPM_{rw} \in ^i_{ov}\mathcal{DCPM}$  with the same o and v as $^iD^o_v$}
\For{$\forall ^im_{qp} \in ^i_{ov}DPM_{rw}$}
\State{$a_p \gets a_p \in ^iD^o_v$ with p from $^im_{qp}$}
\If{there is an $a_{p'} \equiv a_p \in ^{i+1}D^o_{v+1}$}
\State{create new $^im_{q{p'}} = 1$ with q from $^im_{qp}$ and $p'$ from $a_{p'}$}
\State{$^i_{o{v+1}}DPM_{rw} \cup ^im_q{p'}$}
\EndIf
\EndFor
\If{$^i_{o{v+1}}DPM_{rw} \neq 0$}
\State{$^{i+1}\mathfrak{DPM} \cup ^i_{o{v+1}}DPM_{rw}$}
\EndIf
\EndFor
\EndCase
\Case{(4) added $^{i+1}R^r_{w+1}$}
\State{$^{i+1}\mathfrak{DPM} \gets^{i}\mathfrak{DPM}$}
\For{for all mapping sets of previous version: $\forall ^i_{ov}DPM_{rw} \in ^i_{rw}\mathcal{DRPM}$  with the same r and w as $^iR^r_w$}
\State{find equivalent attributes and create new $^im_{qp}$ for new mapping sets and add to  $^{i+1}\mathfrak{DPM}$ similar to case 3}
\EndFor
\State{conduct deletion of previous version as in case 2}
\EndCase
\EndSwitch
\textbf{end switch}
\State{return $^{i+1}\mathfrak{DPM}$}
\EndProcedure
\end{algorithmic}
\label{alg:updatefromdmm}
\end{algorithm}
\subsection{Parallel computation with the DMM}
The full DMM system that is based on partitioned and densely saved permutation matrices enables three-fold parallel computation, namely on the single attribute level, the block level of the Kafka-messages and the system level, as multiple Kafka-message can be processed in parallel within one configuration state $i$ of the system.
\newline
In order to speed up the processing of Kafka-messages from the parallel streams of messages in the streaming pipeline, we can use horizontal scaling of the app. This can be implemented by reading from different Kafka-partitions with different horizontally scaled apps. The DMM-system is horizontally scalable under the condition that we keep the configuration state stable. Thus all scaled apps need to have the same state $i$. Otherwise they may be producing different messages as a result.  
\newline
It is significant to note that we have observed that our configuration state does not change more than a few times a day. Further, we need parallel instances only at rare occasions for defined time-slots, namely for initial loads during which very large numbers of messages need to be processed with several instances that work in parallel. During these slots, changes to the schemata and, therefore, to the distributed system and the matrix, can be disabled.
\newline
Inside a scaled instance, we can conduct the mapping of single messages that we read from the stream with further parallel processes. Each mapping block in one column-super-set of our DMM super-super-set defines a single mapping. We call such a single mapping between two messages an independent mapping path. The great advantage of defining these independent mapping paths in the DMM system consists in the ability to use them for parallel computing.
\newline
Permutation matrices enable a further level of parallel execution. They are consisting of linear independent rows and columns. They, thus, can be deconstructed further to dense sets of single mapping elements with the value 1 for parallel computation. Our hierarchical block-partitioning of the matrix, thus, extends down through the last single atomic mapping element. 
\newline
Finally, we have all elements in place to present the algorithm that executes the mappings with three intertwined processes of parallel computation on a dense set of permutation matrices. It only sends out messages to a parallel stream that contain at least one non-"null" object. For this, we remove the constraint  of the baseline system that demanded that all possible attributes are present in any sparse Kafka-message. We now specify that only attributes with data objects that are not null are present in any dense Kafka-message. Further, we exclude the possibility to send out Kafka-messages with empty payloads. Our JSON-schema enables such compacting to dense messages.
\newline
Since we are operating on dense sets without 0 or "null" values only, we can even simplify the mapping function to a set operation for the DMM. Since all elements $^im_{qp}$ in $^i\mathfrak{DPM}$ have the value 1 and all elements $ad_p$ in an incoming message are non "null", one can say that if we find for a given $ad_p$ a corresponding mapping element with the same index p in the DMM, then it follows that $^im_{qp} = 1$ and $nad_p = 1$. If we plug these values into our mapping function, we obtain 1*1 = 1. Thus, we can create a mapped pair of the attribute $c_q$ and the object $ad_p$, whereby we simple look up the $c_q$ from $^iC$ for the same q as the element $^im_{qp}$ that we have determined before. This simplified mapping function is at the centre of the parallel algorithm.
\newline
Since we can break down the permutation matrices into single atomic mapping operations for parallel computation and can execute parallel computation on the two higher levels of the block-system, too, we have achieved an optimal time efficient solution. As in any parallel architecture, the actual execution time is dependent on the system configuration and the available resources. Thus, we are measuring it later. The algorithm is shown as algorithm \ref{alg:parallel}.
\begin{algorithm}[ht]
\caption{Parallel and dense mapping of $^iDMIn^o_v$ to $^iDMOut^r_w$ with $^i\mathfrak{DPM}$ with set intersection}
\begin{algorithmic}[1]
\Procedure{Map}{Parallel stream of $^iDMIn^o_v$}
\For{$^iDMIn^o_v$ \textbf{in parallel}}
\State{$^i\mathcal{DCPM}^o_v \gets ^i\mathcal{DCPM}^o_v$ from $^i\mathfrak{DPM}$ with the same o and v as in $^iDMIn^o_v$}
\For{$\forall$ $^i_{ov}DPM_{rw} \in ^i\mathcal{DCPM}^o_v$ \textbf{in parallel}}
\State {$^iDMOut^r_w \gets$ create message with empty payload with id of corresponding $^{i}R^r_{w}$}
\For{$ \forall ^im_{qp} \in ^i_{ov}DPM_{rw}$ \textbf{in parallel}}
\If{there is an $^id.s_o.v_v.a_p.ad_p \in ^iMIn^o_v$ for the same index p as $^im_{qp}$}
\State {$ c_q \gets c_q \in ^{i}R^r_{w}$ for the same q as $^im_{pq}$}
\State {payload $^iDMOut^r_w \cup \{c_q, cd_q = ad_p\}$}
\EndIf
\EndFor
\If{payload of $^iDMOut^r_w$ not empty}
\State{send $^iDMOut^r_w$ to \textbf{parallel stream}}
\EndIf
\EndFor
\EndFor
\EndProcedure
\end{algorithmic}
\label{alg:parallel}
\end{algorithm}
\newline
It is further important to note that while all operations occur only once and at once for a single mapping process inside one app, the ETL pipeline does not overall realize an 'exactly once' process. It is possible that FX emits the same data-load twice via different events. This can be identified by unique keys in the payload. Thus, for incoming data events that have a valid mapping, the ETL pipeline with the DMM system ensures an 'at least once' approach. There is also an error-checking and update-process in place  for technically non-valid mappings.
\section {Implementation}
\subsection{A new CDM architecture}
The aim of this paper is to present a new solution for the integration of a large microservice system into a streaming ETL pipeline. As shown above, the position of the CDM mapping system has changed as compared to the established position in an ESB. The consequence of this architectural decision in the case of EOS is that the CDM mapping has been implemented as a microservice within the FX system. Thus, other known  implementations, such as a MapReduce implementation with Hadoop, which are usually conducted in components after the APIs of a microservice system, were not chosen.
\newline
We have, thus, implemented the DMM-system into the microservice METL with the standard stack for microservices that we use at EOS, namely Java Spring-Boot, Postgres, Kafka, and React for the User Interface. In addition, we implemented Debezium  and the Apicurio-Schema-Registry in the new ETL pipeline. The registry is accessible from all stages of FX to enable an early testing of new updates. \cite{Schindler:2021} It contains all the schemata that we use in FX and that map onto the business entities defined in the CDM.
\begin{figure*}[t]
\includegraphics[width=\textwidth]{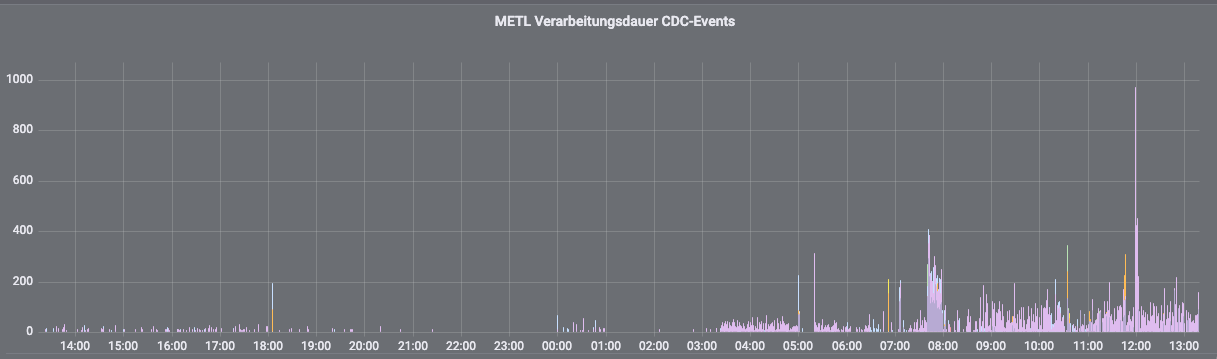}
\caption{METL - Dashboard for system evaluation}
\end{figure*}
\subsection{Implementation of a hybrid system}
We have implemented a hybrid solution that uses both described strategies and, thus, uses two dynamic mapping matrices for different use cases. For the parallel computation, we use the set $^{i}\mathfrak{DPM}$. We keep it in memory. For effective storage in the database, we use the more strongly compacted set $^i\mathfrak{DUSB}$. Since we can recreate $^iM$ from $^i\mathfrak{DUSB}$ and can transform $^iM$ to $^i\mathfrak{DPM}$, we have a clear path to recreate $^i\mathfrak{DPM}$ from $^i\mathfrak{DUSB}$ with two algorithms. The recreation is very helpful if we need to restart the system or copy the configuration to another instance of the app.
\newline
The hybrid solution also changes the update process slightly. 
We save any updates to the matrix to $^i\mathfrak{DUSB}$ directly and then update $^i\mathfrak{DPM}$ from this set. This has allowed us to downshift the update algorithm to a Postgres-View which is based on a SQL-implementation of the decompacting and update algorithm. The compacting to $^i\mathfrak{DUSB}$ after an update helps to recognize new unique permutation matrices about which we inform the user.
\newline
In order to make the access to $^{i}\mathfrak{DPM}$ effective for parallel computing, we use a cached function that reads in the columns $^i\mathcal{DCPM}^o_v$ of $^{i}\mathfrak{DCPM}$ into an efficient hashmap which makes them accessible in O(1). We are using Caffeine as the cache for the function. We evict the cache every time, a business entity, schema or mapping is updated or created and thus enforces an update of the system and the mapping matrix to a new state.
\subsection{User Interface}
Further, we have implemented an User Interface (UI) for enabling a user to create mapping blocks and to confirm updates to a new unique permutation matrix. The confirmation and update procedure is currently dealt with via an error and update process of the mappings, but is scheduled for full automation. The UI, further, enables a detailed inspection of all single mapping paths. The UI provides a good way to enforce the basic rule of the system (as compared to CSV initialisation files), namely the 1:1 attribute mappings that result in permutation matrices, which form the basis of the DMM.
\newline
The main feature request of our data owners has been to implement a reverse search that allows them to see which $^in'$ different Kafka messages with extracting schema versions are mapping to one Kafka message with one business entity version. For this search, we are using the row-set $^i\mathfrak{DRPM}$. Further, the data owners wanted to understand how the version progression is functioning in the system and thus, we have also implemented a search function, which exhibits all mappings with relation to one extracting schema and multiple versions.
\newline
The User Interface has proven to be a good basis to advance the more general evolution model for the versions, also of the business entities. Mappings can be elevated to new business entity versions in the User Interface in the same way as they are extended to new schemata versions. The system is thus currently incrementally upgraded to include more automation steps as outlined in this paper.
\subsection{Reserve capacity}
The system realises a $^im'$:$^in'$ mapping from Kafka-messages with extracting schemata to Kafka-messages with a CDM-schema. We have observed that many extracting schemata versions map to one business entity version only. This is an intended behaviour, as we expect an integration to the CDM to reduce the number of extracting schemata and not to inflate it. Since mappings from one incoming message to several outgoing ones are rare, we currently do not need the second parallel execution process that splits up the mappings to the $^im'$ outgoing messages. However, we have kept it as a reserve capacity to speed up METL further, should the need arise. As outlined above, the same holds true for the usage of horizontal scaling. We currently only need horizontal scaling for an initial load. Thus, also in this respect, we have reserve capacity.
\section{Evaluation}
Finally, we have developed several dashboards that allow for a full monitoring of the application that has gone live now. We record the number of transformations, the time they take  and the storage requirements of the Caffeine cache. Thus, we are able to measure the effectiveness of our solution. 
\newline
METL currently takes on average 39 milliseconds to conduct a full one-to-one-mapping. These data were measured over one day, namely on 13 February 2022, based on 1168 change-data-capture-events  from Debezium. The standard-deviation, however, is high with 51 milliseconds. The two most probable causes for this are: First, METL is hosted in Docker-containers on virtual servers. Thus, external operations can impact the performance of METL. Secondly, the DMM-update is triggered several times a day, which evicts all caches and thus impacts on the processing time of a single event after the eviction. Given the known impact of external factors stemming from the virtual servers, we can conclude that it is reasonable to assume that the standard processing time for a single mapping of one incoming message without cache-update sits very likely in the lower bracket of execution times, which can measured at around 10-20 milliseconds per CDC event.
\section{Conclusion}
In sum, it has been shown that our new DMM-approach presents a highly efficient solution to the overall problem of modernizing ETL pipelines that are based on Kafka-streams and the near real-time extraction of data with CDC events. We were able to solve the most pressing problem of this implementation, namely the usage of a very large and changing mapping matrix with a new type of DMM. Our solution compacts the matrix by more than 99.9\%, conducts the mapping operations with an optimal parallel execution and automates updates. This implementation is a new contribution to various research fields, including streaming ETL pipelines, data warehousing and ML data engineering, enterprise architecture and microservices as well as to the fields of matrix partitioning and dynamic networks. The significance of this innovation derives from the fact that modern ETL pipelines, that extract data from distributed systems and load it to a data warehouse and ML systems, are at the centre of many modern data-driven enterprises.
\section{Authors}
Dr. Christian Haase, Senior Data Scientist, Otto Business Intelligence; Software Engineer, EOS (during the METL-Project); Associated Researcher NLP-Group, Department of Informatics, University of Hamburg
\newline
\newline
Timo Röseler, Senior Software-Architect, EOS, Hamburg, Germany 
\newline
\newline
Dr. Mattias Seidel, Senior Software-Architect, EOS, Hamburg, Germany
\section{Acknowledgements}
For the development of the CDM/ETL-pipeline, EOS assembled a team of system-architects, software engineers and data scientists. Many team-members, among them the authors of this paper, have experience in dealing with very large and fast changing data-sets and complex networks. The group had support from the director of software engineering at EOS-Technological Solutions, as well as from many other engineers, data scientists, requirement engineers, process managers and analysts at EOS and from members of external companies. The authors are indebted to all of them for fruitful discussions, contributions to the code and exchanges of ideas. The solution that we present in this paper would not have been possible without this wider innovative culture at EOS.
\bibliography{anthology,eacl2021}
\bibliographystyle{acl_natbib}
\end{document}